\documentstyle[mhd209]{article}
\input{epsf.sty} 

\title{The $\alpha$ Dynamo Effects in Laboratory Plasmas}

\author{Hantao Ji\inst{1} and Stewart C. Prager\inst{2}}

\institute{Princeton Plasma Physics Laboratory,
Princeton, New Jersey 08543, USA
\and
Department of Physics, University of Wisconsin,
Madison, Wisconsin 53706, USA}

\begin{document}
\maketitle
\begin{abstract}
A concise review of observations of the $\alpha$ dynamo effect in laboratory
plasmas is given. Unlike many astrophysical systems, the laboratory 
pinch plasmas are driven magnetically. When the system is overdriven,
the resultant instabilities cause magnetic and flow fields to 
fluctuate, and their correlation induces electromotive
forces along the mean magnetic field. This $\alpha$-effect drives
mean parallel electric current, which, in turn, modifies the 
initial background mean magnetic structure towards the stable regime.
This drive-and-relax cycle, or the so-called self-organization process, 
happens in magnetized plasmas in a time scale much shorter than resistive 
diffusion time, thus it is a fast and unquenched dynamo process. 
The observed $\alpha$-effect redistributes magnetic 
helicity (a measure of twistedness and knottedness of magnetic field 
lines) but conserves its total value. It can be shown that fast 
and unquenched dynamos
are natural consequences of a driven system where fluctuations are 
statistically either not stationary in time or not homogeneous in 
space, or both. Implications to astrophysical phenomena will be 
discussed.
\end{abstract}

\section{Introduction}

Phenomena involving magnetic field have been observed in astrophysical systems
ranging from planets and stars to accretion disks, galaxies and even in clusters of 
galaxies \cite{Parker79}. Understanding the origins and effects of these cosmic 
magnetic fields has been one of most active research areas across 
multiple subdisplines of physics. In particular, generation and 
sustainment of magnetic fields from dynamics in electrically conducting media, 
or so-called dynamo actions \cite{Moffatt78}, has long remained an unsolved
problem.

A typical astrophysical system is driven by a combination of thermal, 
rotational, and gravitational energies. For example,
dynamics in the outer core of the earth is dominated by Coriolis force 
(due to rotation) and thermal convection (due to temperature gradient and 
gravity). Another example is the accretion disks where the differential
rotation is a primary source of free energy originating from the 
release of gravitational energy. Magnetic field in these systems can 
grow out of corresponding instabilities. Often, the resultant Lorentz force
is ignored in the equation of the motion, i.e., the so-called kinematic dynamo
problem. When the Lorentz force is fully taken into account in the flow 
dynamics, the problem is nonlinear. 

In the kinematic dynamo problem, the flow velocity is completely 
determined by the
corresponding free energy source, either thermal, rotational, or 
gravitational. The growth of magnetic field is only passively determined as a linear 
problem, and is not a part of dynamics. 
The kinematic dynamo introduces simplicity, but does not provide 
a self-consistent solution for either the flow or magnetic field.

The dynamos in astrophysical systems are mostly fully nonlinear 
in nature.  The effects of magnetic field must be fully taken into account 
i.e., the Lorentz force is an integral part of the dynamics, often resulting 
in saturation of the magnetic field growth. In most cases, however, the nonlinear
dynamos are too complicated to be studied theoretically without
invoking statistical treatments. One such example is the so-called
mean-field electrodynamics \cite{Krause80}, where the ensemble-averaged
electromotive force is calculated along the mean magnetic field 
(the $\alpha$-effect) or the mean electric current (the $\beta$-effect).

The mean electromotive force generates and sustains the entire mean magnetic 
field or current, which, in turn, influences the dynamo effects.
This is true in all the astrophysical systems driven non-magnetically.
When a system is driven magnetically, i.e. the free energy source is 
in the magnetic form, the electromotive force or the $\alpha$-effects
still modify the externally supplied mean magnetic field. Sometimes 
the modifications are so significant that the resultant mean field is 
qualitatively different from its initial profile. This article is 
intended to describe such an example in laboratory pinch plasmas 
which are driven only magnetically. The observed $\alpha$-effects 
exhibit remarkably similar properties to those in astrophysical 
systems: they modify the mean magnetic profile in a time-scale
much faster than resistive time scale, and they
are closely related to the concept of magnetic helicity as shall
be described later. The laboratory elucidates key aspects of 
dynamo physics since detailed measurements are possible, 
and the MHD nonlinear problem can be formulated and solved.

The remainder of this paper is organized as follows:
qualitative introduction to laboratory pinch
plasmas is given in Sec.~2, including demonstration of 
the need for a dynamo mechanism to explain magnetic field generation.   
The nonlinear MHD description of the dynamo is summarized in Sec.~3, 
including the role of the dynamo in the self-organization of the plasma, 
and the instabilities that underlie the process. 
Experimental measurements of $\alpha$-dynamo effects are presented
in detail in Sec.~4. The nature of the observed dynamo effects and their
relation to magnetic helicity are discussed in Sec.~5, followed
by implications to the astrophysical dynamos and conclusions in Sec.~6.

\section{Pinch plasmas: magnetically driven systems}

\begin{figure}[t]
\centerline{\epsfxsize=3in\epsfbox{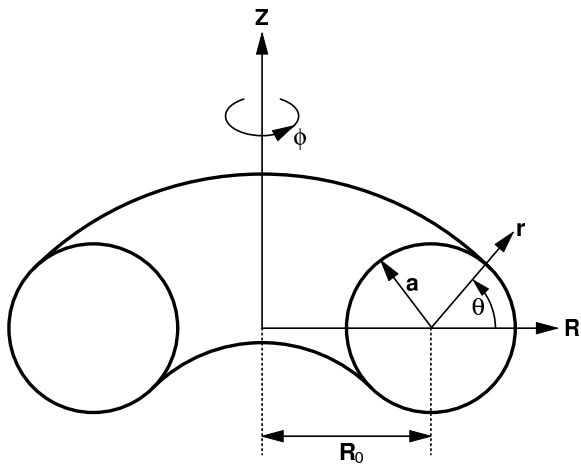}}
\caption{Coordinates for toroidal geometry: the cylindrical coordinate
(major radius $R$, toroidal angle $\phi$, axial 
direction $Z$) and the toroidal coordinate (toroidal
angle $\phi$, poloidal angle $\theta$, and radius $r$). The aspect 
ratio of a torus is defined as $R_0/a$.}
\label{toroidal}
\end{figure}

The history of pinch plasma experiments goes back to early 1950's
as a part of efforts for development of magnetically confined 
plasmas for the nuclear fusion application.
The basis of these experiments is the pinch effect, in which
a current channel in the electrically conducting plasmas contracts
through the self-magnetic field of the current.
The current can be driven either by a voltage applied between electrodes
or electromagnetic induction through a transformer.
We shall focus on a subclass of the pinch 
experiments that are toroidal (Fig.~\ref{toroidal}), 
where most of relevant theoretical and experimental dynamo work 
have been performed. Specifically, the reversed-field pinch (RFP) 
plasmas \cite{Bodin80,Bodin90} shall be described in detail. 
A similar configuration known as spheromak \cite{Jarboe,Bellan}
shall also be mentioned but the interested readers should refer to the 
review papers mentioned above for some experimental details.

\subsection{Formation of pinch plasmas}

\begin{figure}
\centerline{\epsfxsize=3.5in\epsfbox{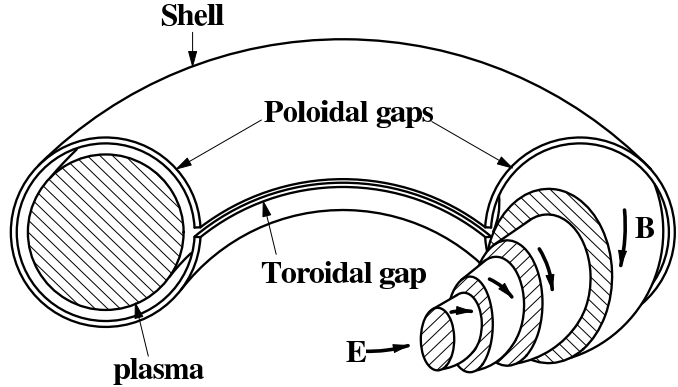}}
\caption{Illustration of an RFP plasma with its magnetic field
at several radii. The toroidal component reverses its direction at
the edge.}
\label{RFP}
\end{figure}

The RFP plasmas are produced by an inductive electric field along the 
toroidal direction through a transformer. The plasma serves
as the secondary coil of the transformer. A typical experimental 
arrangement is conceptually illustrated in Fig.~\ref{RFP}, where the plasma
is surrounded by a metal shell. The electrical skin time of the shell 
is much longer than 
the experimental time thus it can be considered as an ideal conductor to
ensure that the magnetic field lines are always parallel to the shell 
surface. The shell has gaps along toroidal and poloidal directions permitting
instantaneous penetration of the applied electric field. 

\begin{figure}
\centerline{\epsfxsize=4.8in\epsfbox{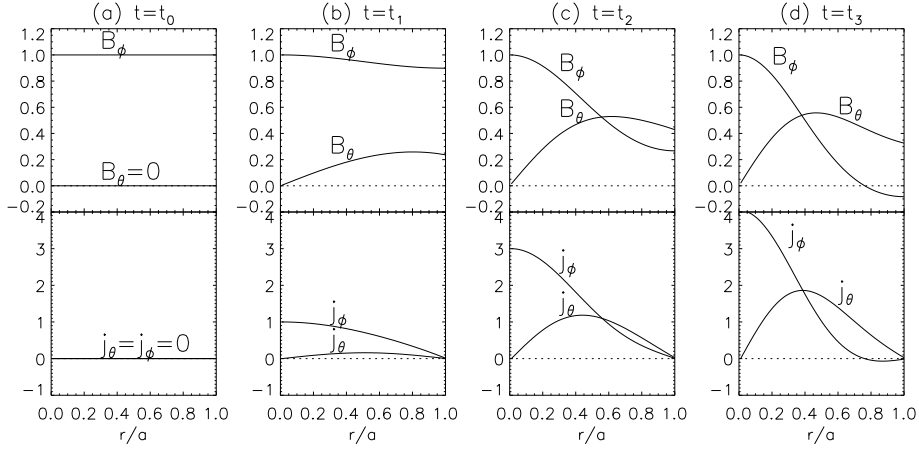}}
\caption{Qualitative illustration of time evolution of the normalized 
magnetic field and current density profiles in a pinch plasma when the electric field is
increased in sequence: (a) initial state where only
toroidal field exists; (b) small electric field to drive mainly 
toroidal current in tokamaks; (c) modest electric field to
significantly drive both toroidal and poloidal current components;
(d) and finally an RFP configuration is realized when the toroidal
magnetic field reverses its direction.}
\label{evolution}
\end{figure}

The time evolution of magnetic field and current density profiles as functions of 
$r$ are qualitatively shown in Fig.~\ref{evolution}.
Initially ($t \leq t_0$), a uniform toroidal magnetic field is imposed 
[Fig.~\ref{evolution}(a)].
At $t=t_0$, the toroidal electric field\footnote{Penetration of
electric field in the conducting plasma is observed to be much faster than
the skin time estimated from the classical resistivity of a plasma. The 
responsible mechanisms have been a subject of research but no clear
answers exist to date.} is applied through the 
transformer to drive the toroidal current, which, in turn, 
produces the poloidal magnetic field. Therefore, the field lines
become helical. Since the electrons freely move along the field lines but 
not so across the field lines, the current path also becomes helical
with a poloidal component. The poloidal current modifies the initially
imposed toroidal field. Figure \ref{evolution}(b) illustrates the
magnetic and current profiles at a time $t=t_1>t_0$, when the 
modification to the initial field is relative small. (This corresponds
to the tokamak configurations, another subclass of toroidal pinch plasmas.) 
We note that the toroidal field increases at the center but 
decreases at the edge.

If the applied toroidal electric field is raised further raised, 
the amplitude of both toroidal and poloidal currents increases
to further modify the field profiles. The poloidal field approaches
to the magnitude of the toroidal field, while the toroidal field 
continues to peak at the center and diminish at the edge, as illustrated in 
Fig.~\ref{evolution}(c) for $t=t_2>t_1$. Finally, when the electric field 
is raised to a large enough value, the toroidal field eventually 
reverses its direction at the edge [Fig.~\ref{evolution}(d) for 
$t=t_3>t_2$,] the origin of the name of reversed-field pinch (RFP).
Figure \ref{RFP} also illustrates magnetic structures at various
radii. Since the 
center toroidal field increases by a much larger value than 
the edge value decreases, 
the total toroidal flux is amplified significantly. 

The spheromak plasma configuration is similar to the RFP.  
It contains helical field lines, with the field at the edge being 
mainly poloidal.  The spheromak is a torus with aspect ratio of 
unity (no hole in the center).  An additional difference from the RFP is 
that the electric field that drives the initial current is mainly 
poloidal and localized at the plasma edge. 
Nonetheless, the observed dynamo effects are strikingly similar to 
those in the RFP. Therefore,
the rest of this paper shall focus on RFP plasmas.

\begin{figure}[b]
\centerline{\epsfxsize=2in\epsfbox{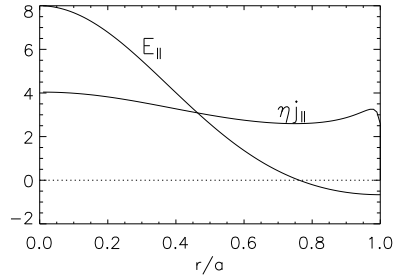}}
\caption{Mismatch between the applied electric field $E_\parallel$ and the resistive 
counterpart $\eta j_\parallel$ along the field line at $t=t_4$ of Fig.~\ref{evolution}(d).
It is noted that $E_\parallel > \eta j_\parallel$ at the center (often called
anti-dynamo) while $E_\parallel < \eta j_\parallel$ at the edge (dynamo).}
\label{e_para}
\end{figure}

\subsection{Need for a dynamo effect}
The current density and magnetic field profile in the RFP, 
particularly the reversal of the toroidal magnetic field, 
cannot arise in a steady-state plasma that is toroidally symmetric 
(lacking in symmetry-breaking fluctuations).
This can be seen easily by 
comparing terms in the MHD Ohm's law parallel to the magnetic field,
\begin{equation}
E_\parallel = \eta j_\parallel.
\label{parallel_ohms_law}
\end{equation}
The parallel component of electric field, $E_\parallel$, can be
calculated by ${\bf E\cdot B}/B=E_\phi B_\phi/B$, where $E_\phi$ is the 
fully penetrated electric field and $B=\sqrt{B_\phi^2+B_\theta^2}$. 
Since both $E_\phi$ and $B$ are a constant, $E_\parallel$ has
the shape of $B_\phi$ which reverses its sign at the edge. On the other hand,
$\eta j_\parallel$ never changes its sign across the radius,
as shown in Fig.~\ref{e_para}. As a result, the edge parallel current,
essentially in the poloidal direction, flows against an externally 
applied electric field. Thus, the parallel component of Ohm's law 
cannot be satisfied without additional terms, such as the 
$\alpha$-effect.

\begin{figure}[t]
\centerline{\epsfxsize=3in\epsfbox{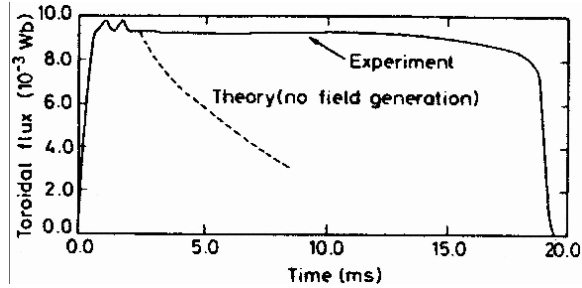}}
\caption{Time evolution of toroidal flux: measurement
and calculation with no dynamo effects in an RFP plasma\cite{Caramana84}.
Note that the decay around $t=18$ms is due to termination of the 
applied electric field.}
\label{toroidal_flux}
\end{figure}

The need for a dynamo effect can also be inferred by comparing the 
measured toroidal flux with that predicted from a simple, symmetric 
resistive MHD theory. 
Figure \ref{toroidal_flux} shows an example of the
measured time evolution of toroidal flux together with
that calculated by resistive MHD theory without dynamo effects.
Interestingly, the toroidal flux is sustained as long as the toroidal 
electric field is provided in the experiment, while the flux decays away 
in the theory. Clearly, existence of dynamo effects is required to 
explain both amplification and sustainment of the toroidal flux.

\section{Self-Organization and the MHD Dynamo}

\begin{figure}[b]
\centerline{\epsfxsize=4in\epsfbox{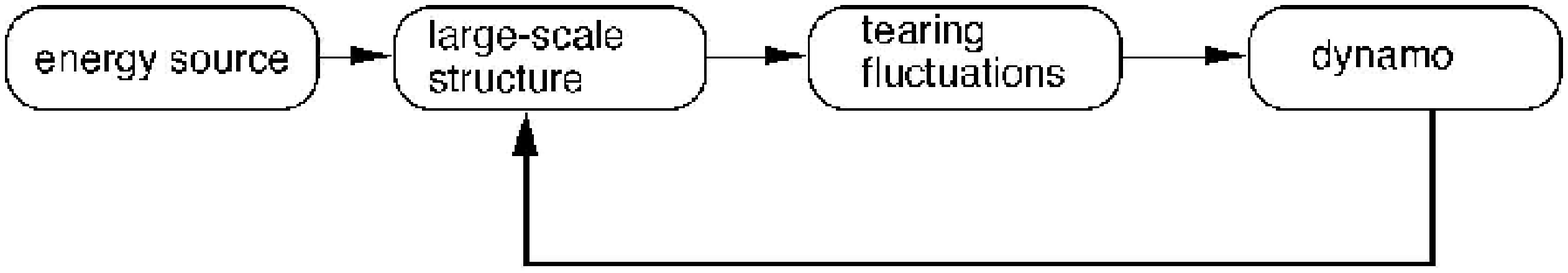}}
\caption{Schematic of magnetic self-organization in laboratory plasmas.}
\label{self_organization}
\end{figure}

The dynamo effect in laboratory plasmas is part of a self-organization 
process. Resistive diffusion evolves the plasma away from a preferred 
state; the dynamo forces the plasma back toward the preferred, 
self-organized state.  The macroscopic features of magnetic 
self-organization are described in Sec.~3.1.  
Resistive diffusion is well-understood; 
the large scale-self-organization driven by the dynamo 
is a current topic of research and the focus of this paper.  
Single-fluid MHD equations provide a fully nonlinear 
description of the dynamo in laboratory plasmas.   
The fluctuations, in velocity and magnetic field, and the 
large-scale mean magnetic field, are all determined self-consistently.  
The MHD calculations include both the effect of the fluctuations 
on the mean field and the effect of the mean field on the fluctuations.  
Weakly nonlinear analytic calculations capture some of the key physics, 
and computational solution of the nonlinear, three-dimensional resistive 
MHD equations provide a complete description. The fluctuations that 
underlie the dynamo arise from tearing instabilities. A brief discussion 
of tearing instabilities in the laboratory context is included in 
Sec.~3.2.  A description of some of the key results of the nonlinear 
problem is presented in Sec.~3.3. 

\subsection{Magnetic self-organization}
The dynamo effect underlies magnetic self-organization in the 
laboratory~\cite{Ortolani93}. 
The plasma is driven by an applied electric field as seen in 
Sec.~2.1.  
The resulting magnetic configuration has excess free energy 
leading to instabilities (or turbulence) that relax the plasma 
toward a state of lower magnetic energy.  The plasma relaxes to 
the lower energy state through the dynamo effect. The self-generation 
of plasma current reconfigures the large-scale magnetic field to one 
with lower energy.  This process is depicted in 
Fig.~\ref{self_organization}.

\begin{figure}
\centerline{\epsfxsize=3in\epsfbox{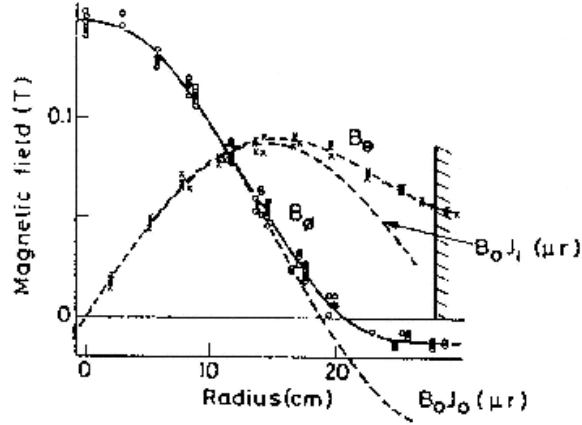}}
\caption{Solution for axial and azimuthal magnetic field of the 
relaxed state and its comparisons with the measurements~\cite{Bodin90}.}
\label{Bessel}
\end{figure}

The structure of the relaxed state is partly captured by minimizing 
the magnetic energy in the plasma volume ($W=\int(B^2/2\mu_0)dV$) 
subject to the constraint of constant magnetic helicity 
($K=\int {\bf A}\cdot {\bf B}dV$), 
where ${\bf A}$ is the magnetic vector potential). Magnetic helicity 
is a topological measure of the knottedness of the magnetic field 
lines~\cite{Taylor86}.  
The minimization yields a magnetic field given by $\nabla \times {\bf B} 
= \mu {\bf B}$, 
where $\mu$ is a constant.  Thus, the ratio of the current density 
to magnetic field, $j/B$, is a spatial constant.  
This relaxed state is sometimes referred to as the Taylor state.  
For a one-dimensional cylindrical plasma (where ${\bf B} = {\bf B}(r)$) 
the solution yields Bessel functions, $B_z = B_0J_0(\mu r)$, $B_\theta 
= B_0J_1(\mu r)$, 
where $B_0$ is a constant.  The noteworthy feature of this solution, 
shown in Fig.~\ref{Bessel}, is that it approximates the measured profiles of 
the reversed field pinch shown in Fig.~\ref{evolution}d 
(identifying $B_z$ with the 
toroidal field and $B_\theta$ with the poloidal field).  
In particular, the reversal of the toroidal field is obtained.  

\begin{figure}
\centerline{\epsfxsize=4in\epsfbox{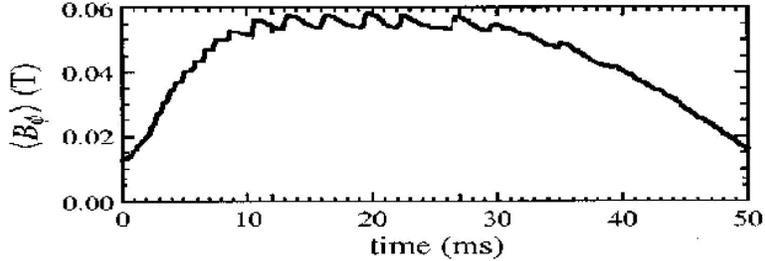}}
\caption{Toroidal magnetic flux versus time in an RFP experiment.}
\label{sawtooth}
\end{figure}

The driven/relaxation phenomenon can occur in experiment 
in either a continuous or cyclic fashion.  The drive forces the 
plasma away from the relaxed state, and the dynamo opposes this 
tendency.  In some experimental plasmas, the net result is that 
these effects are nearly balanced at all times; thus, the plasma 
mean field is approximately steady in time.  In other experimental 
plasmas, the two effects are separated in time~cite{Watt83,Hokin91}.  
During the drive period, the plasma slowly evolves away from the 
relaxed state, driven by the applied electric field.  
During the relaxation period, the dynamo rapidly returns to 
the plasma to a relaxed.  This sawtooth cycle is evident 
experimentally in the toroidal magnetic flux in the plasma, 
shown in Fig.~\ref{sawtooth}.  A slow decay of the flux is followed by rapid 
magnetic flux generation by the dynamo.  The dynamo acts as discrete 
events in time.  During the decay phase the ratio $j/B$ is becoming 
spatially peaked; during the relaxation phase it becomes flatter. 
Relaxation processes happen also in spheromak plasmas~\cite{Yamada99}
where the driving electric field can be either in toroidal or poloidal
direction with the aspect ratio, $R_0/a$, close to unity, but the
underlying physics of relaxation is essentially the same.

\begin{figure}[b]
\centerline{\epsfxsize=3in\epsfbox{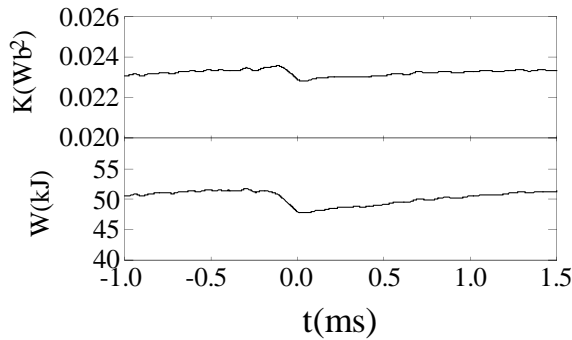}}
\caption{Time evolution of total magnetic helicity and energy 
during relaxation event in an RFP plasma~\cite{Ji95a}.}
\label{MST_helicity}
\end{figure}

An experimental test of the Taylor conjecture was achieved 
by inferring the change in magnetic energy and helicity 
during a relaxation event~\cite{Ji95a}.  An approximate measurement was obtained 
by modeling the instantaneous plasma state as a slowly varying 
MHD equilibrium.  Since the fields during relaxation are changing 
on a time scale longer than an Alfven time, the fields will 
satisfy the MHD force balance equation, ${\bf j} \times {\bf B} 
= \nabla p$, through a 
discrete dynamo event.  Through solution of this equation with 
experimental constraints it is found that during a relaxation 
event the magnetic energy reduces by about 8\%, while the 
magnetic helicity reduces by about 3\%, as shown in 
Fig.~\ref{MST_helicity}. (Also see Sec.~5.4)

\subsection{Magnetic (tearing) instabilities}

The Taylor conjecture provides a useful framework to depict 
approximately the final state of the relaxation, or dynamo, 
process. However, it provides no information on the physical 
mechanism of the dynamo.  Solution of the resistive nonlinear 
MHD equations reveals the detailed dynamics.  The spatial 
fluctuations that underlie the dynamo are tearing instabilities.  
Tearing instabilities are driven by spatial gradients in $j/B$, and 
cause the field lines to tear and reconnect~\cite{Furth63}.  The MHD description 
of such spontaneous magnetic reconnection has been investigated 
for several decades.  From linear theory, it is known that tearing 
instabilities grow on a timescale that is intermediate between 
short Alfven time ($a/V_A$, where 
$V_A = (B^2/\mu_0 \rho)^{1/2}$ and $\rho$ is the mass density) 
and the long resistive 
diffusion time ($\mu_0 a^2/\eta$, where $\eta$ is the resistivity).  
For the experimental plasma parameters, the theoretical 
growth time is approximately 100$\mu s$, comparable 
the observed fast relaxation time.

The tearing instability can be described by a wave function of the 
form $f(r)\exp[i(m\theta  - n\phi)]$ where m and n are integers representing 
the poloidal and toroidal mode numbers.  In laboratory plasmas, 
many tearing modes can be present simultaneously.  
For tearing to occur, the fluctuating field must be constant 
along the mean magnetic field.  That is, the parallel wave 
number must vanish (parallel wavelength must become infinite) 
somewhere in the plasma.  This condition ($k_\parallel = 0$) becomes
\begin{equation}
{\bf k}\cdot {\bf B} =  k_\theta B_\theta + k_\phi B_\phi =
{m\over r}B_\theta - {n \over R} B_\phi = 0.
\label{tearing}
\end{equation}
The condition can be 
written as $q = m/n$ where $q = rB_\phi/RB_\theta$ is the winding number 
of the mean magnetic field and varies with minor radius.  This condition 
represents a resonance between the fluctuating and mean magnetic 
fields.  In laboratory plasma discussed here, multiple tearing 
modes with multiple mode numbers occur. Hence, tearing occurs 
at many radii within the plasma, leading to large-scale reorganization.

\subsection{The nonlinear MHD dynamo}

An $\alpha$-effect dynamo arises from MHD tearing instabilities, 
as indicated in the mean-field Ohm¹s law
\begin{equation}
    \overline {\bf E} + \langle \widetilde {\bf v} \times
    \widetilde {\bf B} \rangle = \eta \overline {\bf j}
\label{MHDohm}
\end{equation}
where the tilde denotes fluctuations and $\langle~\rangle$ 
denotes mean quantities 
(averages over the poloidal and toroidal directions).   
Some key features of the dynamo electromotive force that 
arises from the fluctuations can be discerned from quasilinear 
theory.  In quasilinear theory the alpha effect term 
(the second term on the left hand side) is evaluated from the 
solutions for the fluctuations obtained from linear stability theory. 
In linear theory the mean field is specified; the spatial structure and 
the growth rate of the exponentially growing modes are calculated. 
From quasilinear theory the parallel component of the alpha effect 
in the vicinity of the radius about which reconnection occurs is 
found to be~\cite{Strauss85,Bhattacharjee86}
\begin{equation}
    \langle \widetilde {\bf v} \times
    \widetilde {\bf B} \rangle_\parallel = 
    \nabla \cdot \left( D \nabla {\overline j \over B} \right)
\label{quasilinear}
\end{equation}
This indicates that the dynamo effect drives current so as to 
reduce the gradient in $j/B$, consistent with a relaxation in the 
direction of the Taylor state.  The dynamo is in the form of a 
diffusion mechanism. The diffusion coefficient is proportional to
$\widetilde B^2$. 

\begin{figure}
\centerline{\epsfxsize=2.5in\epsfbox{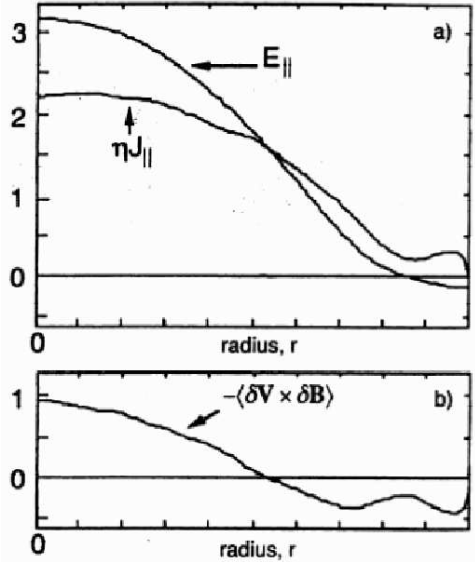}}
\caption{Radial profile of each term in the mean-field Ohm's law,
showing (a) the mean electric field and current density and
(b) the fluctuation-induced dynamo term. Each term is evaluated from
nonlinear MHD computation.}
\label{simulation}
\end{figure}

Quasilinear theory is incomplete. It only captures the dynamo 
effect during the growing phase of the instability.  
Extensive computational study has produced a fully self-consistent description 
of the MHD laboratory dynamo.  The nonlinear computation 
predicts a steady state dynamo, and includes the interaction 
between the fluctuations and mean fields (the quasilinear effect) 
and the nonlinear energy transfer between different spatial Fourier 
modes.  Computation predicts that the fluctuation energy is spread 
over a modest number of nonlinearly interacting modes, of order of 
the aspect ratio, $R_0/a$. 
From an initial state of random noise, the instabilities grow 
(excited by the gradient in $j/B$), reaching an amplitude of about 
1\% of the mean field.  The radial profile of each of the terms 
in the parallel (to $\overline {\bf B}$) component of the mean-field Ohm's law 
is displayed in Fig.~\ref{simulation}.  Note that the alpha effect is large.  
At the radius where the mean parallel electric field is zero, 
all the current is driven by the dynamo.  Since the alpha effect 
reverses sign with radius, it can also be viewed as redistributing 
current from the center the edge. 

\section{Measurements of Dynamo Effects}

Despite the long history of research of RFP plasmas and subsequent
recognition of existence of dynamo effects, the experimental
efforts to directly measure the $\alpha$-effect did not start until
the late 1980's \cite{Ji92}. The main reason is the 
difficulty in measuring local velocity fluctuations of both ions and electrons 
in a plasma. Below we describe several techniques successfully used to measure
the velocity fluctuations, thus the dynamo effects.
In addition to the MHD dynamo, discussed in the previous section, 
measurement have been made to examine dynamo effects arising from 
pressure effects, the Hall term, and kinetic effects. 
We discuss each of these four dynamo mechanisms in turn.

\subsection{MHD dynamo}
The key quantity to measure is the component of the fluctuation-induced
electromotive force along the mean magnetic field. In the MHD model, 
where ${\bf v}$ is dominated by the ${\bf E} \times {\bf B}$ drift,
the electron and ion velocities are about equal.  
The corresponding term is $\langle \widetilde {\bf v} 
\times \widetilde {\bf B} \rangle_\parallel \simeq 
\langle \widetilde {\bf v}_\perp \times \widetilde {\bf B}_\perp
\rangle_\parallel$.
Thus, the key is to measure $\widetilde {\bf v}_\perp$ with sufficient 
time and spatial resolutions.

The first technique invokes measurements of fluctuations in the perpendicular 
electric field. The measuring principle is based on the fluctuating Ohm's law,
$\widetilde {\bf E} + \widetilde {\bf v} \times \overline
{\bf B} + \overline {\bf v} \times \widetilde {\bf B}= \eta \widetilde{\bf j}$, 
which reduces to
\begin{equation}
\widetilde {\bf v}_\perp = {(\widetilde {\bf E}_\perp + 
\overline {\bf v} \times \widetilde {\bf B}
-\eta \widetilde{\bf j}_\perp) \times \overline {\bf B} \over \overline B^2}
\simeq  {\widetilde {\bf E}_\perp \times \overline {\bf B} \over \overline B^2},
\label{perp_ohms_law}
\end{equation}
because $\widetilde {\bf E}_\perp \gg \eta \widetilde {\bf j}_\perp$ and
$\widetilde {\bf E}_\perp \gg \overline {\bf v} \times \widetilde {\bf B}$
in typical conditions. Therefore, the $\alpha$-effect becomes
\begin{equation}
\langle \widetilde {\bf v} \times \widetilde {\bf B} \rangle_\parallel \simeq 
{\langle ( \widetilde {\bf E}_\perp \times \overline {\bf B} ) \times 
\widetilde {\bf B}_\perp \rangle \over \overline B^2}
= {\langle \widetilde {\bf E}_\perp \cdot \widetilde {\bf B}_\perp \rangle \over 
\overline B}.
\label{edotb}
\end{equation}

Fluctuations in electric and magnetic fields and their correlations 
can be measured by Langmuir probes and magnetic pickup coils, respectively,
at the plasma edge where the temperature is relatively low \cite{Ji91}. 
However, the first attempt using this technique in a relatively dense 
RFP plasma measured \cite{Ji92} no perceivable $\alpha$-effect compared 
to the mismatch between $E_\parallel$ and $\eta j_\parallel$ shown in 
Fig.~\ref{e_para}. It has been suggested later that another mechanism
(see next subsection) for the dynamo action might be operational in
these dense plasmas.

\begin{figure}
\centerline{\epsfxsize=4in\epsfbox{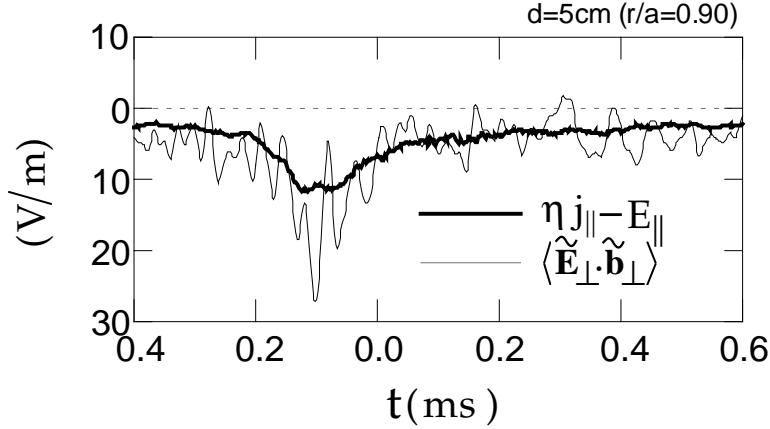}}
\caption{Time evolution of the measured MHD dynamo electric field 
$\langle \widetilde {\bf E}_\perp \cdot \widetilde {\bf B}_\perp \rangle / 
\overline B$ by Langmuir probes \cite{Ji94} in MST, compared with the mismatch 
$\eta j_\parallel - E_\parallel$
in the parallel Ohm's law during a relaxation cycle.}
\label{MST_Langmuir}
\end{figure}

The first successful detection \cite{Ji94} of the MHD dynamo in the RFP plasmas
was made in the well-controlled Madison Symmetric Torus (MST) plasmas.
As discussed in Sec.~3.1, in addition to the continuous dynamo action to
regenerate toroidal flux against resistive diffusion, discrete
relaxation events occur in MST in a regular fashion to generate toroidal 
flux in a short time scale. Figure \ref{MST_Langmuir} displays the 
measured dynamo electric field compared with the mismatch between 
$E_\parallel$ and $\eta j_\parallel$.
The agreement is excellent both during relaxation events ($t \simeq -0.1$ms) 
and between events.
Successful measurements of MHD dynamo
effects were made in a spheromak using a similar technique \cite{Alkarkhy93}.

\begin{figure}
\centerline{\epsfxsize=4.5in\epsfbox{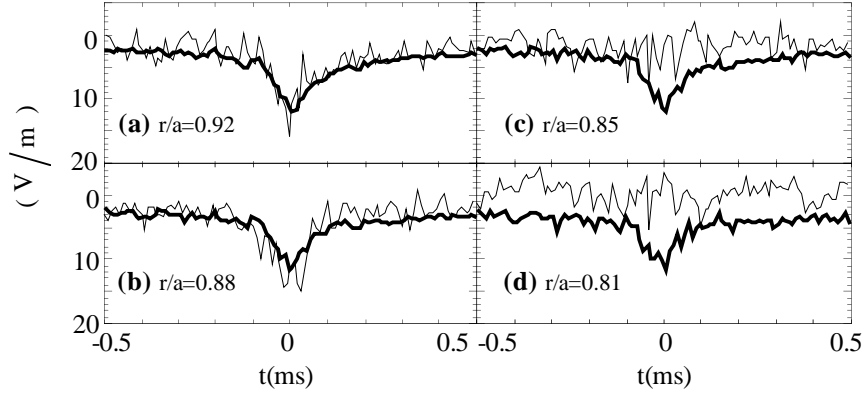}}
\caption{Time evolution of the measured MHD dynamo electric field 
$\langle \widetilde {\bf v} \times \widetilde {\bf B} \rangle_\parallel$ 
(light line) by an optical probe \cite{Fontana00} in MST, compared with the 
mismatch $\eta j_\parallel - E_\parallel$ (heavy line)
in the parallel Ohm's law during a relaxation cycle.}
\label{MST_IDSP}
\end{figure}

The second technique used to measure fluctuating velocities is 
spectroscopic detection of the Doppler shift from impurity carbon ions
embedded within the hydrogen plasmas \cite{DenHartog98,Fontana00}.
If the coupling between impurity ions and plasma ions are strong, or
they behave in the same way when exposed to the slow electric field
fluctuations, the emission from impurity ions function as a tracer
of plasma flow. An optical probe \cite{Fiksel98} was utilized to measure
local velocity fluctuations of impurity ions. Correlations with
local magnetic fluctuations have also yielded remarkable agreement
with the mismatch electric field at the edge as shown in 
Fig.~\ref{MST_IDSP}. At smaller radii, however, the measured
dynamo term diminishes due to the phase changes in $\widetilde v_r$
\cite{Fontana00}. Other mechanisms discussed in Sec.~3 
for the dynamo effects might be operational at the smaller radii.

\subsection{Diamagnetic dynamo}
Because of the collisionless nature of high temperature plasmas,
MHD approximations are not always a good model to describe dynamics
of such plasmas. A better model can be based on the two-fluid model
where ion fluid and electron fluid are treated separately. As 
described in Sec.~3.3, the force balance for electrons is 
essentially the Ohm's law, which is generalized to include
electron pressure force,
\begin{equation}
{\bf E} + {\bf v}_e \times {\bf B} + {\nabla P_e \over en} = \eta {\bf j},
\label{twofluid}
\end{equation}
where the electron inertial effect is ignored.
Then Eq.(\ref{edotb}) is modified to
\begin{equation}
\langle \widetilde {\bf v}_e \times \widetilde {\bf B} \rangle_\parallel \simeq 
{\langle \widetilde {\bf E}_\perp \cdot \widetilde {\bf B}_\perp \rangle \over 
\overline B}
+{\langle {\bf \nabla}_\perp \widetilde P_e \cdot \widetilde {\bf B}_\perp \rangle 
\over e \overline n \overline B}
\label{diamagnetic}
\end{equation}
where the second term is referred as ``diamagnetic'' dynamo since the 
fluctuating electron velocity is due to electron diamagnetism.
We note that a small ``battery'' term is ignored here (see Sec.~5.1.)

\begin{figure}[t]
\centerline{\epsfxsize=5in\epsfbox{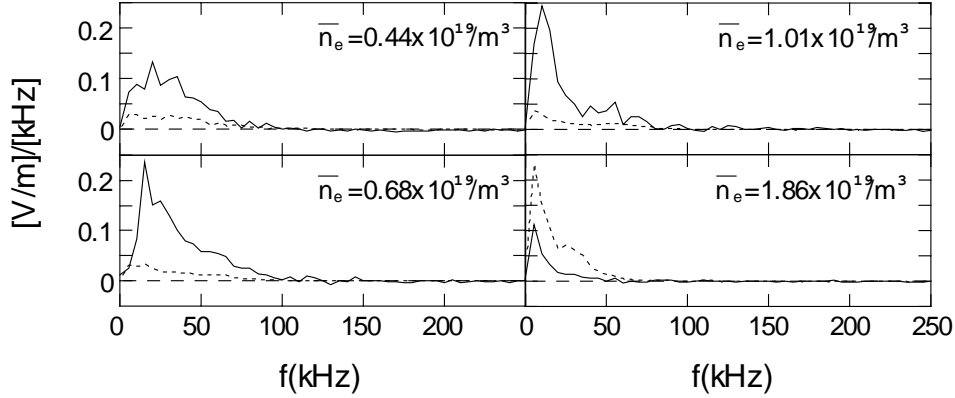}}
\caption{Cross spectra of the MHD (solid lines) and diamagnetic dynamo
(dotted lines) terms for four different density cases in a RFP plasma 
\cite{Ji95b}.}
\label{TPE}
\end{figure}

The diamagnetic dynamo has been detected at the edge of an RFP plasma when
the density is high \cite{Ji95b}. Fluctuations in the electron pressure
(density and electron temperature) can be measured by Langmuir probes
at different locations to deduce their gradient. Figure~\ref{TPE} shows the 
cross spectra of fluctuating quantities for both MHD dynamo and 
diamagnetic dynamo for four different densities.
In the low density plasmas, the MHD dynamo dominates 
while at the highest density, the diamagnetic dynamo dominates.
Although the underlying mechanisms for the transition are still unclear,
this observation is consistent \cite{Ji96} with the early measurements 
in a dense plasma where no significant MHD dynamo was detected \cite{Ji92}. 

\subsection{Hall dynamo}
Since ${\bf j} = en ({\bf v}_i -{\bf v}_e)$ and ${\bf v} \approx {\bf 
v}_i$, Eq.(\ref{twofluid}) can be written as
\begin{equation}
{\bf E} + {\bf v} \times {\bf B} - {{\bf j} \times {\bf B} \over en} 
+{\nabla P_e \over en}= \eta {\bf j},
\label{Hall}
\end{equation}
where the second term in the left-hand side is called the Hall term.
The fluctuation-induced counterpart, $\langle \widetilde {\bf j} 
\times \widetilde {\bf B} \rangle /en$, therefore called Hall dynamo,
has been suggested to be important under certain conditions \cite{Nebel90}.
Experimentally, current density fluctuations can be measured by
magnetic pickup coils, placed at several spatial points, using Ampere's law.
Measurements at the plasma edge indicated small but non-negligible
effects in the Ohm's law \cite{Shen93}.

Physically, the Hall dynamo arises when electron flow does not 
fluctuate together with ions in the perpendicular direction. 
For this to happen, electrons need to experience different
forces than ions. Electric force cannot be the cause since its
fluctuations induce only the same flow fluctuations for both species.
However, the electron pressure force can serve this purpose by
driving flows only in electrons. Therefore, the diamagnetic dynamo 
mentioned above is a manifestation of the Hall dynamo effect.
Another possibility for a finite difference in perpendicular flows of 
electrons and ions has been suggested \cite{Ji95b,Nebel90}
when the viscous force for ions is large.

\subsection{Kinetic dynamo}
\begin{figure}
\centerline{\epsfxsize=3.5in\epsfbox{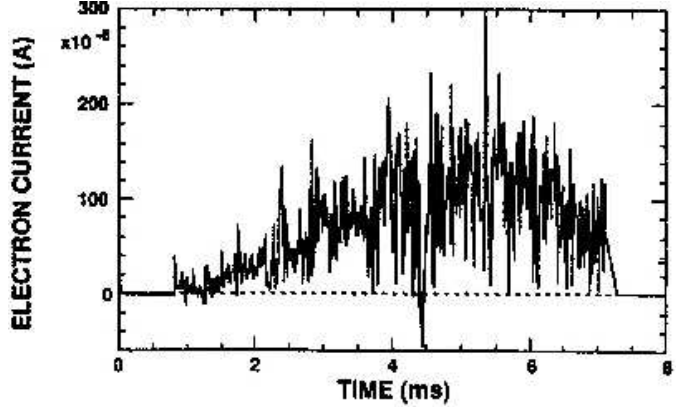}}
\caption{Time evolution of energetic electron current
measured by an electrostatic electron energy analyzer
with the entrance aligned along the field line
at the edge of an RFP plasma \cite{Ingraham90}.}
\label{fast_e}
\end{figure}

For ions and electrons to be treated as fluids, their 
distribution functions need to be close to Maxwellians. In the 
collisionless plasmas, however, this cannot be always true
especially in the pinch plasmas driven by a large electric field.
In such cases, often a high-energy tail of electrons form
along magnetic field line. When the field lines wander from center
to edge, these high-energy electrons follow, 
resulting in transport of electric current outward.
The current transport due to these
electrons can just make up the mismatch in the electric field
illustrated in Fig.~\ref{e_para}. The physics process, however,
needs to be treated by the full kinetic equations, referred
as the kinetic dynamo theory \cite{Jacobson84}. The main support
for this theory is from the observation \cite{Ingraham90}
of high energy electrons along the field lines carrying most 
of the current, as exemplified in Fig.~\ref{fast_e}.

However, the present form of this theory does not include
the effects of these energetic electrons on the dynamics
of magnetic field lines through Ampere's law. It has been pointed
out \cite{Terry90} that such dynamical feedback to the magnetic field structure
would impose a severe constraint on the efficiency of this mechanism.
Alternately, the observation of energetic electrons is possibly 
just a manifestation of other dynamo processes motioned above,
i.e., they are accelerated by local dynamo electric field
\cite{Yagi99}. Kinetic behaviors of the plasma in such a turbulent 
environment may be far from a simple physics picture.

\section{Implications to Dynamo Theories and Astrophysical Dynamos}

What we can learn from the observed dynamo effects described in the
previous sections? In this section, we discuss their implications to
a number of important and often controversial issues in recent 
developments in dynamo theory: fast dynamo versus slow dynamo;
back-reaction of mean field on dynamo action;
stationary and homogeneous turbulence versus driven systems;
and its relationship with magnetic helicity.
Finally, implications to astrophysical systems, especially
the solar dynamo problem, are discussed.

\subsection{Fast dynamo versus slow dynamo}

Fast dynamos are dynamo actions which do not diminish in the small
resistivity limit, while slow dynamos do. Fast dynamos can change
the magnetic field in a much faster time scale than the resistive
diffusion time, in which slow dynamos operate. A real solution for the
astrophysically observed dynamos must be a fast dynamo.
In fact, the dynamo effects described in the previous section 
are evidently also fast dynamos, which can change the magnetic flux in 
a time scale (typically $\sim 100 \mu$s during relaxation events) much 
faster than the resistive diffusion time (typically $\sim 0.1$s).  

For clarity, we re-derive the $\alpha$-effect based on the
Ohm's law in the two-fluid framework, Eq.~(\ref{twofluid}),
${\bf E} + {\bf v}_e \times {\bf B} + \nabla P_e/en = \eta {\bf j}$.
After ensemble-averaging over fluctuations it becomes
\begin{equation}
\overline {\bf E} + \overline {\bf v}_e \times \overline {\bf B} +
\langle \widetilde {\bf v}_e \times \widetilde {\bf B} \rangle
+ {\nabla \overline P_e \over e n} = \eta \overline {\bf j},
\label{ohm_mean}
\end{equation}
which can be subtracted from Eq.~(\ref{twofluid}) to yield
\begin{equation}
\widetilde {\bf E} + \widetilde {\bf v}_e \times \overline {\bf B}
+ (\overline {\bf v}_e + \widetilde {\bf v}_e) \times \widetilde {\bf B}
- \langle \widetilde {\bf v}_e \times \widetilde {\bf B} \rangle
+ {\nabla \widetilde P_e \over en} 
= \eta \widetilde {\bf j},
\label{ohm_turb}
\end{equation}
where small battery-like effects such as 
$\langle \widetilde n \nabla \widetilde P_e\rangle /e\overline n^2$ 
are neglected (see Sec.~5.4). With the use of Eq.~(\ref{ohm_turb}),
the $\alpha$-effect is calculated as
\begin{eqnarray}
\langle \widetilde {\bf v}_e \times \widetilde {\bf B} 
\rangle_\parallel  & = & {\langle \widetilde {\bf v}_e \times \widetilde {\bf B} 
\rangle \cdot \overline {\bf B} \over B} =
- {\langle (\widetilde {\bf v}_e \times \overline {\bf B})
\cdot \widetilde {\bf B} \rangle \over B}
\nonumber \\
& = & {\langle \widetilde {\bf E} \cdot \widetilde {\bf B}\rangle 
\over B}
+{\langle \nabla \widetilde P_e \cdot \widetilde {\bf B} \rangle 
\over e \overline n B}
-\eta {\langle \widetilde {\bf j} \cdot \widetilde {\bf B}\rangle 
\over B}.
\label{dynamo}
\end{eqnarray}
The first and second terms are MHD dynamo and diamagnetic dynamo, 
respectively, and the third term is proportional to the resistivity
$\eta$. We note that the so-called
``first-order smoothing'' approximation \cite{Krause80},
or $\widetilde {\bf v}_e \times \widetilde {\bf B}
- \langle \widetilde {\bf v}_e \times \widetilde {\bf B} \rangle =0$
needs not to be assumed in Eq.~(\ref{ohm_turb}) to derive the above results.

The third term in Eq.~(\ref{dynamo}) deserves special comment.
It is likely a slow dynamo term since it diminishes as the resistivity goes
to zero. It may be argued that the current density fluctuations may go
to infinity to make this term finite in the vanishing resistivity 
limit~\cite{Seehafer95}.
But this possible singularity is a mathematical question but not a physical
one because current density has to be bounded in any case:
microinstabilities will be destabilized sooner or later to stop the 
growth in the current density. Therefore, this term should be always 
a slow dynamo term in real plasmas. Indeed, this term is small in  
pinch plasmas where resistivity is small, as mentioned in the previous 
section.
In contrast, the other two terms in Eq.~(\ref{dynamo}) can be
the fast dynamo terms since they are not constrained by the small
resistivity, and they can change the magnetic field in a fast time 
scale, as demonstrated in the laboratory plasmas described 
in this paper. We discuss these terms further in the following subsections.

\subsection{Back-reaction of mean field on dynamo action}

The concept of slow dynamo is closely related, if not identical,
to the so-called back-reaction of mean magnetic field on the $\alpha$-
effect in the analytic MHD models 
\cite{Kulsrud92,Gruzinov94,Bhattacharjee95,Vainshtein98},
confirmed by MHD simulations \cite{Cattaneo96}.
A self-consistent constraint due to Lorentz force on the flow
\cite{Pouquet76} applies on the kinematic dynamo effects,
\begin{equation}
\alpha \equiv {\langle \widetilde {\bf v} \times \widetilde {\bf B} 
\rangle_\parallel \over B}
=\alpha_0+{\tau \over 3 \rho} \langle \widetilde {\bf j} \cdot
\widetilde {\bf B} \rangle,
\label{alpha}
\end{equation}
where the kinematic $\alpha$-effect $\alpha_0 = -(\tau /3 \rho)
\langle \widetilde {\bf v} \cdot \nabla \times \widetilde {\bf v} \rangle$
and $\tau$ is the correlation time same for
the velocity and magnetic field fluctuations. Combining with
Eq.~(\ref{dynamo}) in the MHD limit,
\begin{equation}
\alpha = {\langle \widetilde {\bf E} \cdot \widetilde {\bf B} \rangle
-\eta \langle \widetilde {\bf j} \cdot \widetilde {\bf B} \rangle 
\over B^2},
\label{MHDdynamo}
\end{equation}
yields \cite{Bhattacharjee95}
\begin{equation}
\alpha = {\alpha_0 + {\tau \over 3 \rho \eta} 
\langle \widetilde {\bf E} \cdot \widetilde {\bf B} 
\rangle \over 1 + {\tau \over 3 \rho \eta} B^2}.
\label{constraint}
\end{equation}
In the limit of small resistivity, the above equation becomes
\begin{equation}
\alpha = {3 \rho \eta \over \tau B^2} \alpha_0 +
{\langle \widetilde {\bf E} \cdot \widetilde {\bf B} 
\rangle \over B^2}
\label{limit}
\end{equation}
where, without the second term, the $\alpha$-effect diminishes either by
small resistivity (or slow dynamo) or by large mean field, $B$ 
(back-reaction of mean field) \cite{Gruzinov94}. In either case, 
however, the second term can survive \cite{Prager99}
to function as a fast, unquenched dynamo. 
In fact, the dynamo effects described in the previous section operate 
in the pinch plasmas with small resistivity in a strong magnetic 
field (although part of it is supplied externally);
thus they are not quenched through the backreaction of the strong
magnetic field. An important question to be discussed in the 
following subsection is under what conditions the fast dynamo terms 
can survive to dominate the dynamo 
effect in systems with small resistivity and/or large mean magnetic 
field. 

\subsection{Stationary and homogeneous turbulence versus driven systems}

First, let us turn our attention to the classical case of statistically
stationary and homogeneous turbulence \cite{Krause80}. In this 
special case, by definition, all statistical quantities of the
turbulence do not vary in time and space. Therefore, the second term
of Eq.~(\ref{dynamo}), $\langle \nabla \widetilde P_e \cdot \widetilde
{\bf B} \rangle = \nabla \cdot \langle \widetilde P_e \widetilde {\bf B}
\rangle$, vanishes. 
Since $\widetilde {\bf E}=-\nabla \widetilde \phi - \partial \widetilde 
{\bf A}/\partial t$, where $\phi$ is the electrostatic potential and ${\bf A}$
is the vector potential\footnote{Theoretically, gauge-invariant treatments of 
these potentials are often not so straightforward, but experimentally
both parts of the electric field can be measured unambiguously:
a double Langmuir probe, consisting of two small electrodes
in contact with plasma, can detect $-\nabla \phi$ along the direction
linking two electrodes while a loop wire with a straight part exposed to
electromagnetic induction but with the rest shielded can detect
$-\partial {\bf A} / \partial t$ along the direction of the straight 
part \cite{Stenzel79}.}, the first term of Eq.~(\ref{dynamo}) 
becomes $\langle \widetilde  {\bf E} \cdot \widetilde {\bf B} \rangle 
= -\langle \nabla \widetilde \phi \cdot \widetilde {\bf B} \rangle - 
\langle \partial (\widetilde {\bf A} / \partial t) \cdot \widetilde {\bf B}
\rangle$. Obviously, the electrostatic part, $-\langle \nabla \widetilde 
\phi \cdot \widetilde {\bf B} \rangle= -\nabla \cdot \langle \widetilde \phi
\widetilde {\bf B}\rangle$, vanishes. Using vector identities and 
${\bf B} = \nabla \times {\bf A}$, the electromagnetic part vanishes
as well:
\begin{equation}
-\left< {\partial \widetilde {\bf A} \over \partial t}
\cdot \widetilde {\bf B}\right>
=-{1\over 2} \left( {\partial \langle \widetilde {\bf A} \cdot
\widetilde {\bf B} \rangle \over \partial t} + 
\nabla \cdot \left< \widetilde {\bf A} \times {\partial \widetilde
{\bf A} \over \partial t}\right> \right) = 0.
\label{K_t}
\end{equation}
Therefore, in the case of stationary and homogeneous turbulence, the 
only surviving dynamo term in Eq.~(\ref{dynamo}) 
is $\eta \langle \widetilde {\bf j} \cdot \widetilde {\bf B} \rangle$, 
which is likely slow, and consequently, the dynamo is quenched in 
Eq.~(\ref{limit}). This is exactly what was predicted theoretically 
\cite{Kulsrud92,Gruzinov94,Bhattacharjee95,Vainshtein98}.
When periodic boundary conditions are imposed, MHD simulations
\cite{Cattaneo96,Brandenburg01a} can produce such turbulence, 
demonstrating slow and quenched dynamos \cite{Blackman00a}.

Stationarity and homogeneity of a turbulence implies that
there exist no preferential directions in time and space
for the system to evolve. This is not the case for the pinch 
plasmas as we have seen so far. They are driven systems. 
In fact, almost all astrophysical systems are also driven by various
sources, as discussed in Sec.~1. More importantly, there always exist
preferences in directions in time and/or in space during the driven 
processes. Often, when the system evolves very slowly so that 
quasi-stationarity can be assumed, the homogeneity in space 
still cannot be assumed. The inhomogeneous nature of the driving
processes can be reflected in the boundary conditions \cite{Blackman00a}.
In fact, MHD simulations with open boundary conditions 
\cite{Brandenburg96,Brandenburg01b}
exhibit fast growth of large scale magnetic field,
on a time scale intermediate between the fast Alfven wave transit time
(or eddy turnover time) and the slow resistive diffusion time.

Since stationary and homogeneous turbulence produces 
only slow and quenched dynamos, a logical next question then
is under what conditions the turbulence can drive fast and
unquenched dynamos? In other words, what effects do the first two
terms of Eq.~(\ref{dynamo}) have on the turbulence? Answering this
question leads to the relation of dynamo effects with magnetic helicity.

\subsection{Magnetic helicity and dynamo effects}

As discussed in Sec.~3, magnetic helicity is relatively conserved
compared to magnetic energy during relaxation where dynamo 
effects play important roles. The dynamo effects conserve 
the total magnetic helicity except for resistive effects and a small 
battery effect \cite{Ji99}. This can be seen easily from its rate of change,
\begin{equation}
{d K \over dt } = - 2 \int {\bf E} \cdot {\bf B} dV 
- \int ( 2 \phi {\bf B} + {\bf A} \times {\partial {\bf A} \over \partial t}) 
\cdot d{\bf S},
\label{balance_K}
\end{equation}
where $V$ is enclosed by the surface ${\bf S}$.
Using Eq.~(\ref{twofluid}), the first term becomes
\begin{equation}
\int {\bf E} \cdot {\bf B} dV = \int \eta {\bf j} \cdot {\bf B} dV
+ \int {\nabla P_e \cdot {\bf B} \over en} dV.
\label{conserve}
\end{equation}
The first term on the RHS is a resistive effect, 
which vanishes with zero resistivity.
The second term requires finite pressure gradient, especially 
electron temperature, along the field line to change the total helicity.
However, we note that such parallel gradients are very small owing to 
fast electron flow along the field lines. Such effects, 
often called the battery 
effect~\cite{Parker79}, provide only a seed for magnetic 
field to grow in a dynamo process and, of course, it can be accompanied 
by small but finite magnetic helicity. 
However, the change in $K$ observed during a relaxation event~\cite{Ji95a},
as shown in Fig.~\ref{MST_helicity}, is larger than estimated changes by
resistive and battery effects.

There are two ways for (the fast) dynamo effects to conserve total magnetic
helicity: transport helicity across space~\cite{Boozer86,Hameiri87}
[also see Eq.~(\ref{quasilinear}) which can be written into a surface 
term] or convert it to a different (often larger) scale, the so-called inverse 
cascading~\cite{Pouquet76,Stribling90}. 
Defining helicity in the mean field $K_m \equiv
\int \overline {\bf A} \cdot \overline {\bf B} dV$ and in the 
fluctuations $K_t \equiv \int \widetilde {\bf A} \cdot \widetilde {\bf B} dV$,
using Eqs.~(\ref{ohm_mean}-\ref{dynamo}) and Eq.~(\ref{balance_K})
in some algebra including cancellations and rearrangements of terms yields~\cite{Ji99}
\begin{eqnarray}
{d K_m \over dt } &=&
- 2 \int \left(\eta \overline {\bf j} \cdot \overline {\bf B}
+ \eta \langle \widetilde {\bf j} \cdot \widetilde {\bf B}\rangle 
+ \left<  {\partial \widetilde {\bf A} \over \partial t}
     \cdot \widetilde {\bf B}\right> \right) dV \nonumber \\
& & - \int \left( 2 \bar \phi \overline {\bf B}
        -2{\overline P_e \overline {\bf B} \over en}
        +\overline {\bf A} \times {\partial \overline {\bf A} 
	\over \partial t} + 2\langle \widetilde \phi \widetilde {\bf B}\rangle
	-2{\langle \widetilde P_e \widetilde {\bf B} \rangle \over en}\right) \cdot 
d{\bf S} \label{mean} \\
{d K_t \over dt } &=& 
2 \int \left<  {\partial \widetilde {\bf A} \over \partial t} 
\cdot \widetilde {\bf B}\right> dV 
- \int \left< \widetilde {\bf A} \times {\partial \widetilde {\bf A} \over
        \partial t}\right> \cdot d{\bf S},
\label{turb}
\end{eqnarray}
where the last equation is identical to Eq.~(\ref{K_t}).
The resistive slow dynamo term, appearing as the second term of 
Eq.~(\ref{mean}), changes the mean helicity also in the resistive time scale.
The electrostatic MHD dynamo and diamagnetic dynamo,
both appearing in the surface integral, transport the mean helicity 
across space while the inductive part of the MHD dynamo $-\langle 
(\partial \widetilde {\bf A}/\partial t) \cdot \widetilde {\bf 
B}\rangle$, appearing in the volume 
integral in the both equations but with opposite signs, converts helicity from 
the fluctuations to the mean field~\cite{Ji99}.
Other terms in the surface integral represent various ways to inject
or extract helicity from the volume~\cite{Ji99}. For instance, the third term 
in the surface integral of Eq.~(\ref{mean}) indicates the helicity 
injection by the transformer in RFP plasmas. Since measurements in RFP 
plasmas indicate that the inductive electric field fluctuations 
are much smaller than their electrostatic counterpart by at least one
order of magnitude, the fast dynamo effects observed in 
the pinch plasmas are accompanied by a helicity flow, which is disallowed
in the stationary and homogeneous turbulence. 

The flow of magnetic helicity due to the fast dynamo effects have
been directly measured in RFP plasmas. For
magnetic helicity to be physically meaningful, 
a gauge-invariant definition~\cite{Bevir81} needs to be
used for the double-connected toroidal plasmas bounded by a conducting shell
as illustrated in Fig.~\ref{RFP}:
$ K_{\rm toroidal} \equiv \int {\bf A} \cdot {\bf B} dV 
- \Phi_\phi(a)\Phi_\theta(a) $
where $\Phi_\phi(a)$ is the total toroidal flux and
$\Phi_\theta(a)$ is the poloidal flux threading the central hole of the 
toroidal plasma. Therefore, the rate of change for $K_{\rm toroidal}$ 
is given by
\begin{equation}
    {dK_{\rm toroidal} \over dt} = - 2\int {\bf E} \cdot 
{\bf B} dV + 2\Phi_\phi(a)V_\phi(a),
\label{Kbalance}
\end{equation}
where $V_\phi(a)$ is the toroidal loop voltage at the plasma surface
$r=a$. The plasma can be divided into two parts: a core part at $0\leq 
r\leq b$ and an edge part at $b\leq r \leq a$. Then the total helicity
is the sum of three parts:
core helicity, $K_{\rm core}$,
edge helicity, $K_{\rm edge}$,
and the single linkage between edge poloidal flux 
and core toroidal flux, $K_{\rm link}$.
The balance equation for $K_{\rm edge}$ and $K_{\rm link}$ 
can be written as~\cite{Ji95a}
\begin{equation}
{dK_{\rm edge} \over dt} 
+{dK_{\rm link} \over dt} = 
- 2\int_b^a \eta {\bf j} \cdot {\bf B}dV + 
2\Phi_\phi(a)V_\phi(a)-2\Phi_\phi(b)V_\phi(b) 
+ 2\int \langle \widetilde \phi \widetilde B_r \rangle dS_b,
\label{local}
\end{equation}
where $S_b$ is the surface area at $r=b$ and $V_\phi(b)$
is the toroidal loop voltage at $r=b$. 
The last term represents helicity transport across $r=b$ by
correlation between fluctuations in electrostatic potential $\widetilde \phi$
and radial field $\widetilde B_r$ associated with the MHD dynamo effect.
(The helicity flux due to diamagnetic dynamo was small in this 
case.)

\begin{figure}
\centerline{\epsfxsize=3in\epsfbox{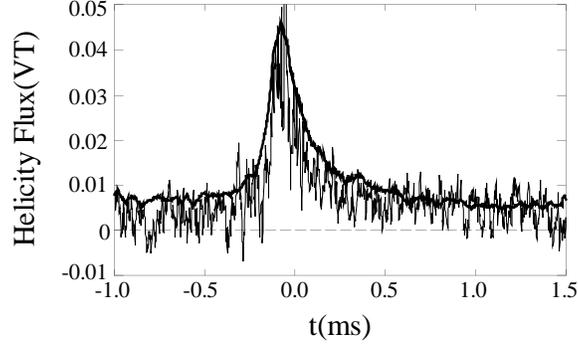}}
\caption{Time evolution of measured helicity flux (thin line) due to MHD
dynamo effects compared to the prediction (thick line) from the helicity balance
during a relaxation event~\cite{Ji95a}.}
\label{helicity_flux}
\end{figure}

The helicity flux has been measured directly for the same plasma shown 
in Fig.~\ref{MST_Langmuir} at $r=b \equiv a-5$cm. All other
five terms in Eq.~(\ref{local}) were determined independently. As shown in 
Fig.~\ref{helicity_flux}, the measured outward helicity flux due to MHD 
dynamo can exactly account for helicity change in $K_{\rm edge}$ and
$K_{\rm link}$, demonstrating the existence of a helicity flux
associated with fast dynamos observed in RFP plasmas. We note that 
the quasi-steady state of the plasma is maintained by an inward helicity
injection by the transformer via the third term of the surface integral of
Eq.~(\ref{mean}).

The requirement of helicity flow in a driven system 
for fast and unquenched dynamos has 
important implications to the physics of astrophysical 
dynamos~\cite{Blackman00b,Vishniac01}.
A good example under debate is the solar dynamo, which is driven by a 
combination of thermal gradient and rotation.
It has been found that there is a preference in the sign of the observed 
twisted field lines (hence the helicity)
in each hemisphere~\cite{Rust94a,Pevtsov95} and consequently 
in the solar wind~\cite{Rust94b}. 
This helicity preference may well be a result of helicity flow accompanied
with the solar dynamo, which must be fast and unquenched 
to explain the observed 11 year solar cycle. 
If the electrostatic MHD dynamo is operational like in laboratory plasmas,
there are two possible explanations. The first possibility~\cite{Ji99} is that
the fast dynamo actions transport or separate the large-scale helicity 
of one sign to one hemisphere while leaving the opposite helicity in the 
other hemisphere. Then they rise to the solar surface via buoyancy.
The second possibility~\cite{Blackman00b} is that the fast dynamo produces
a helicity flow from the convection zone to the surface to drive
the corona. The sign of the helicity is different at each hemisphere 
due to opposite signs of the $\alpha$ effect.
Both mechanisms conserve the total helicity. 
These large-scale structures and its associated helicity
are constantly removed from the solar surface by flaring.
Both mechanisms can also replace the lost helicity continuously.

\section{Conclusions}

The physics of the $\alpha$ effects measured in the pinch plasmas has 
been reviewed, largely based on simple, intuitive approaches. It has 
been demonstrated that the observed dynamo effects are part of 
magnetic self-organization processes, either continuous or discrete 
in time.  MHD dynamo and diamagnetic dynamo have been successfully
measured and there is supporting evidence on Hall dynamo
and kinetic dynamo. The close relationship with the concept of magnetic 
helicity has been studied, and it has been directly measured that the
dynamo activity is accompanied by an outward helicity flow.

Implications of these measurements and understanding have been
discussed in the context of astrophysical dynamos.
Despite important differences in the form of the driving free energy, 
dynamos in laboratory plasmas exhibit remarkable similarities with 
the required astrophysical dynamos:
they are fast (operational at small resistivity) and unquenched
(operational in a large mean field). It has been shown that these desired 
features cannot exist in the traditionally assumed stationary and homogeneous 
turbulence. In fact, both the laboratory plasmas and astrophysical dynamos 
are driven systems which invalidate the stationarity and/or homogeneity 
assumption on the generated turbulence, thus allowing fast and 
unquenched dynamos to function. 
Despite several caveats, the 
availability of laboratory plasmas provides a unique and useful
testbed to enhance understanding of astrophysical problems,
such as dynamo.

%\section*{Acknowledgments}

\newcommand{\noopsort}[1]{}

\received{\today}
%\clearpage

\end{document}